\documentclass[unnumsec,webpdf,contemporary,large]{oup-authoring-template}%

\usepackage{color, colortbl}

\graphicspath{{Fig/}}

\theoremstyle{thmstyleone}%
\theoremstyle{thmstyletwo}%
\theoremstyle{thmstylethree}%

\begin{document}

\journaltitle{arXiv}
\DOI{Preprint}
\copyrightyear{2021}
\pubyear{2021}
\access{OmiTrans}
\appnotes{Problem Solving Protocol}

\firstpage{1}


\title[OmiTrans]{OmiTrans: generative adversarial networks based omics-to-omics translation framework}

\author[1]{Xiaoyu Zhang}
\author[1,2]{Yike Guo}

\authormark{Zhang et al.}

\address[1]{\orgdiv{Data Science Institute}, \orgname{Imperial College London}, \orgaddress{\postcode{SW7 2AZ}, \state{London}, \country{UK}}}
\address[2]{\orgdiv{Department of Computer Science}, \orgname{Hong Kong Baptist University}, \orgaddress{\state{Hong Kong}, \country{China}}}

\corresp[$\ast$]{Corresponding author. \href{x.zhang18@imperial.ac.uk}{x.zhang18@imperial.ac.uk}}

\received{Date}{0}{Year}
\revised{Date}{0}{Year}
\accepted{Date}{0}{Year}



\abstract{With the rapid development of high-throughput experimental technologies, different types of omics (e.g., genomics, epigenomics, transcriptomics, proteomics, and metabolomics) data can be produced from clinical samples. The correlations between different omics types attracts a lot of research interest, whereas the stduy on genome-wide omcis data translation (i.e, generation and prediction of one type of omics data from another type of omics data) is almost blank. Generative adversarial networks and the variants are one of the most state-of-the-art deep learning technologies, which have shown great success in image-to-image translation, text-to-image translation, etc. Here we proposed OmiTrans, a deep learning framework adopted the idea of generative adversarial networks to achieve omics-to-omics translation with promising results. OmiTrans was able to faithfully reconstruct gene expression profiles from DNA methylation data with high accuracy and great model generalisation, as demonstrated in the experiments.} 



\maketitle

\section{Introduction}
The research on multi-omics data is an emerging topic that attracts a lot of interests recently \citep{Wang2021MOGONETIM,Withnell2021XOmiVAEAI,Zhang2021OmiEmbedAU,Efremova2020ComputationalMF}, and the correlation between different omics types, especially the regulation of gene expression, is an classical research topic with long-term development \citep{Jaenisch2003EpigeneticRO, Roundtree2017DynamicRM, Fernandes2019LongNR}. Nevertheless, the genome-wide generation and prediction of one type of omics data (e.g., gene expression) from another type of omics data (e.g. DNA methylation), a.k.a, omics-to-omics translation, has not been well studied. \cite{Zhong2019PredictingGE} was one of the few work that discussed the possibility of predicting gene expression using DNA methylation data, which adopted locally-connected LASSO regression to the task. However, they came up with the conclusion that DNA methylation data had limited prediction power for gene expression, which is mainly because of the rather low capacity of their network according to our experiments. TDimpute \citep{Zhou2020ImputingMR} is a more recent work in this field that applied a quite straightforward neural network to impute missing RNA-Seq data from DNA methylation data. They claimed that TDimpute significantly outperformed other state-of-the-art methods including singular value decomposition imputation (SVD), trans-omics block missing data imputation (TOBMI) \citep{Dong2019TOBMITB}, and LASSO \citep{Tibshirani1996RegressionSA}. TDimpute \citep{Zhou2020ImputingMR} is currently the most state-of-the-art method that is able to predicting gene expression from DNA methylation, and it is our major comparing method.

Generative adversarial networks (GANs) \citep{Goodfellow2014GenerativeAN} are one of the most emerging deep learning methodologies which have seen rapid improvement these years \citep{Karras2017ProgressiveGO, Brock2018LargeSG, Donahue2019LargeSA}. GANs and the variants \citep{Denton2015DeepGI, Salimans2016ImprovedTF, Isola2017ImagetoImageTW}, especially the idea of conditional GAN, have been adapted to solve tasks like image-to-image translation \citep{Isola2017ImagetoImageTW, Zhu2017UnpairedIT, Yi2017DualGANUD} and text-to-image translation \citep{Xu2018AttnGANFT, Reed2016GenerativeAT, Qiao2019MirrorGANLT, Cheng2020RiFeGANRF}. Recently, there are also some attempts to adopt GANs to omics data, especially the omics data imputation task \citep{Xu2020scIGANsSR, Marouf2020RealisticIS}. However, there is current no generative adversarial networks framework designed for the omics-to-omics translation task.

In this work, we proposed OmiTrans, a deep learning framework that adopted the idea of generative adversarial networks (GANs) to achieve omics-to-omics translation. With OmiTrans we were able to faithfully reconstruct gene expression profiles from the corresponding DNA methylation data with high accuracy, which outperformed other methods including the state-of-the-art TDimpute \citep{Zhou2020ImputingMR}. We also applied the pre-trained OmiTrans model to individual and previously unseen datasets which indicated the great applicability, practicability, and model generalisation of OmiTrans.

\section{Materials and methods}
\subsection{Datasets}
The multi-omics pan-cancer database The Cancer Genome Atlas (TCGA) \citep{Weinstein2013TheCG} was selected for the following experiments since TCGA is one of the most comprehensive databases containing different types of omics data for corresponding samples. Although OmiTrans is capable for facilitating the data conversion between any two omics type, we chose DNA methylation and RNA-Seq gene expression profiling in TCGA to demonstrate the omics type conversion ability of OmiTrans. Therefore, we selected 9,081 samples in TCGA with both DNA methylation and gene expression profiles for further experiments. The multi-omics data of TCGA were downloaded from the UCSC Xena data portal (\url{https://xenabrowser.net/datapages/}, accessed on 1 May 2019). 

For RNA-Seq gene expression data, all of the 60,483 probes from UCSC Xena were kept as the the fragments per kilobase of transcript per million mapped reads (FPKM) format to quantify the gene expression level. For DNA methylation data, the original profiles comprise 485,577 probes from the Infinium HumanMethylation450 BeadChip (450K) array. To balance the dimensionality and reduce the computational requirements, feature filtering was applied to the DNA methylation data. The feature filtering is a two-step process. First, certain probes in the 450K array were removed according to the similar criteria mentioned in \cite{Zhang2021OmiEmbedAU}: probes excluded in the current Infinium MethylationEPIC BeadChip (EPIC) array (n = 32,260), probes containing the single-nucleotide polymorphism (SNP) dbSNP132Common within five base pairs of the targeted CpG site (n = 7,998),  probes not uniquely mapping to the human reference genome (hg19) with one mismatch allowed (n = 3,965), the non-CpG loci probes (n = 3,091), the SNP assay probes (n = 65), and probes with missing values (N/A) in more than 10\% of samples (n = 2). Unlike the filtering criteria in \cite{Zhang2021OmiEmbedAU}, we kept the CpG sites in sex chromosomes because those features were required to reconstruct the corresponding gene expression values in the same chromosome. In total, 46,330 probes were filtered in the first step. Then in the next step, the variance of each remaining CpG site was calculated and sorted in descending order. Probes with variance lower than 0.05 were removed for further experiments. Eventually 39,464 CpG sites in the 450K array were remained after the two-step feature filtering process. Any remaining N/A values in DNA methylation and gene expression datasets were replaced by the average feature value. Location alignment is the final preprocessing step, where the feature orders of both the DNA methylation and gene expression profiles were rearranged according to their locations in the chromosomes and the chromosomes were sorted by the order from 1 to 22 and X then Y.

To further evaluate the model generalisation of OmiTrans, we chose the brain tumour DNA methylation dataset (BTM) GSE109381 \citep{Capper2018DNAMC} as a individual dataset to test the performance of model trained by the TCGA dataset. The GSE109381 BTM dataset is one of the largest DNA methylation datasets for brain tumour samples from the Gene Expression Omnibus (GEO). The raw data generated from 450K arrays were downloaded directed from GEO (\url{https://www.ncbi.nlm.nih.gov/geo/query/acc.cgi?acc=GSE109381}, accessed on 30 August 2019), and then beta values that indicate the DNA methylation percentage of each CpG site were calculated by the Bioconductor R package minfi \citep{Aryee2014MinfiAF}. The subsequent preprocessing steps for the DNA methylation data in the BTM dataset were the same as those have been done for the TCGA dataset, and the two DNA methylation datasets shared the same CpG feature list with the same order.

\subsection{Overview of OmiTrans}
\begin{figure}[t!]
    \includegraphics[width = 1\hsize]{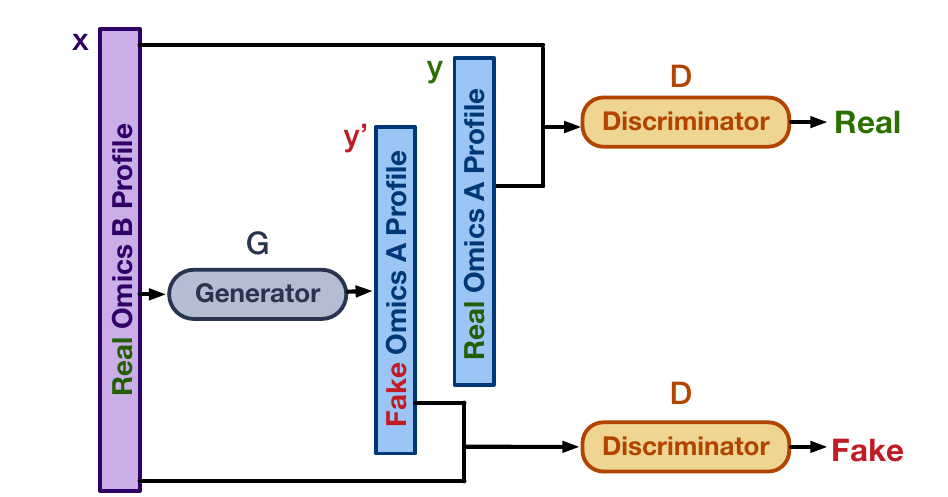}
    \centering
    \caption{The overview of the OmiTrans omics type conversion framework. The main objective of OmiTrans is to train a conditional GAN to map profiles of one omics type (omics type B) to corresponding profiles of another omics type (omics type A). The generator $G$ learns to synthesise omics A profiles, and the discriminator $D$ learns to identify real omics pair from the fake (synthetic) omics pair.}
    \label{fig:CGAN}
\end{figure}

The OmiTrans deep generative framework was proposed to implement the data conversion (or translation) between any two omics types (e.g., gene expression, DNA methylation, and miRNA expression) based on the idea of generative adversarial networks (GANs) and its variants \citep{Goodfellow2014GenerativeAN, Denton2015DeepGI, Salimans2016ImprovedTF, Isola2017ImagetoImageTW}. In a nutshell, OmiTrans learns a mapping from a profile of one omics type $x$ and a random noise vector $z$, to a profile of another omics type $y$, $G:\{x, z\} \rightarrow y$. The discriminator $D$ is trained to classify between the synthetic (fake) $\{x,y\prime\}$ omics pairs and the original (real) $\{x,y\}$ omics pairs. The overall diagram of the OmiTrans framework was illustrated in Figure \ref{fig:CGAN}.

In the original version of GANs \citep{Goodfellow2014GenerativeAN}, the generator $G$ produces the fake data $y\prime$ only from the noise $z$ and the discriminator $D$ classifies between the real data and fake data without the observation of $x$. The value function $V(G,D)$ of a vanilla GAN can be expressed as
\begin{equation}
    \begin{aligned}
    V_{GAN}(G, D)=& \mathbb{E}_{y}[\log D(y)]+\\
    & \mathbb{E}_{z}[\log (1-D(G(z))].
    \end{aligned}
    \label{eq:vfGAN}
\end{equation}
Moving one step forward, we can provide $x$ to generator $G$ to facilitate the generation of $y\prime$ based on corresponding $x$. The value function $V(G,D)$ can be therefore written as
\begin{equation}
    \begin{aligned}
    V_{GAN}(G, D)=& \mathbb{E}_{y}[\log D(y)]+\\
    & \mathbb{E}_{x,z}[\log (1-D(G(x,z))].
    \end{aligned}
    \label{eq:vfcGAN1}
\end{equation}
By also conditioning $x$ in the discriminator $D$, we can get the value function in its complete expression:
\begin{equation}
    \begin{aligned}
    V_{GAN}(G, D)=& \mathbb{E}_{x,y}[\log D(x,y)]+\\
    & \mathbb{E}_{x,z}[\log (1-D(x,G(x,z))].
    \end{aligned}
    \label{eq:vfcGAN2}
\end{equation}
where the generator $G$ tries to minimise the function to fool the discriminator $D$, whereas the discriminator $D$ tries to maximise the function to spot the fake data generated by $G$, i.e., 
\begin{equation}
    \arg \min _{G} \max _{D} V_{GAN}(G,D).
    \label{eq:minmax}
\end{equation}

To make the synthetic omics data $y\prime$ more related to the corresponding $x$, we also added a distance constrain to the objective of OmiTrans. The generator $G$ tries to produce the synthetic omics data $y\prime$ that not only fool the discriminator $D$ but be as similar as possible to the original omics data $y$, and the value function of the distance constrain is
\begin{equation}
    V_{dist}(G)=\mathbb{E}_{x,y,z}[dist(y-G(x,z))]
    \label{eq:dist}
\end{equation}
In the OmiTrans framework, we provide L1, mean square error (MSE) and cross entropy (CE) as the distance function $dist()$. With the distance constrain, the final objective of OmiTrans can be formalised as
\begin{equation}
    \arg \min _{G} \max _{D} V_{GAN}(G,D)+\lambda V_{dist}(G)
    \label{eq:finalobj}
\end{equation}
where $\lambda$ balances the GAN loss and the distance loss.

\subsection{Network Architecture}
The OmiTrans framework supports various implementations of generator $G$ and discriminator $D$. For the similar image translation task, most solutions \citep{Isola2017ImagetoImageTW,Radford2015UnsupervisedRL,Wang2016GenerativeIM,Yoo2016PixelLevelDT,Zhu2016GenerativeVM} applied deep convolutional layers to implement $G$ and $D$ because the characteristics of image data fit the sparse connectivity and weights sharing of convolutional neural networks (CNNs), and the underlying structure of the input image is aligned with the structure of the output image. However it is not applicable to omics data because the format of omics data is not 2D/3D gird where each pixel/voxel is connected to its neighbours normally with similar values, and the input omics data and the output data with different dimensionality do not share the same underlying structure. 

\begin{figure}[t!]
    \includegraphics[width = 0.9\hsize]{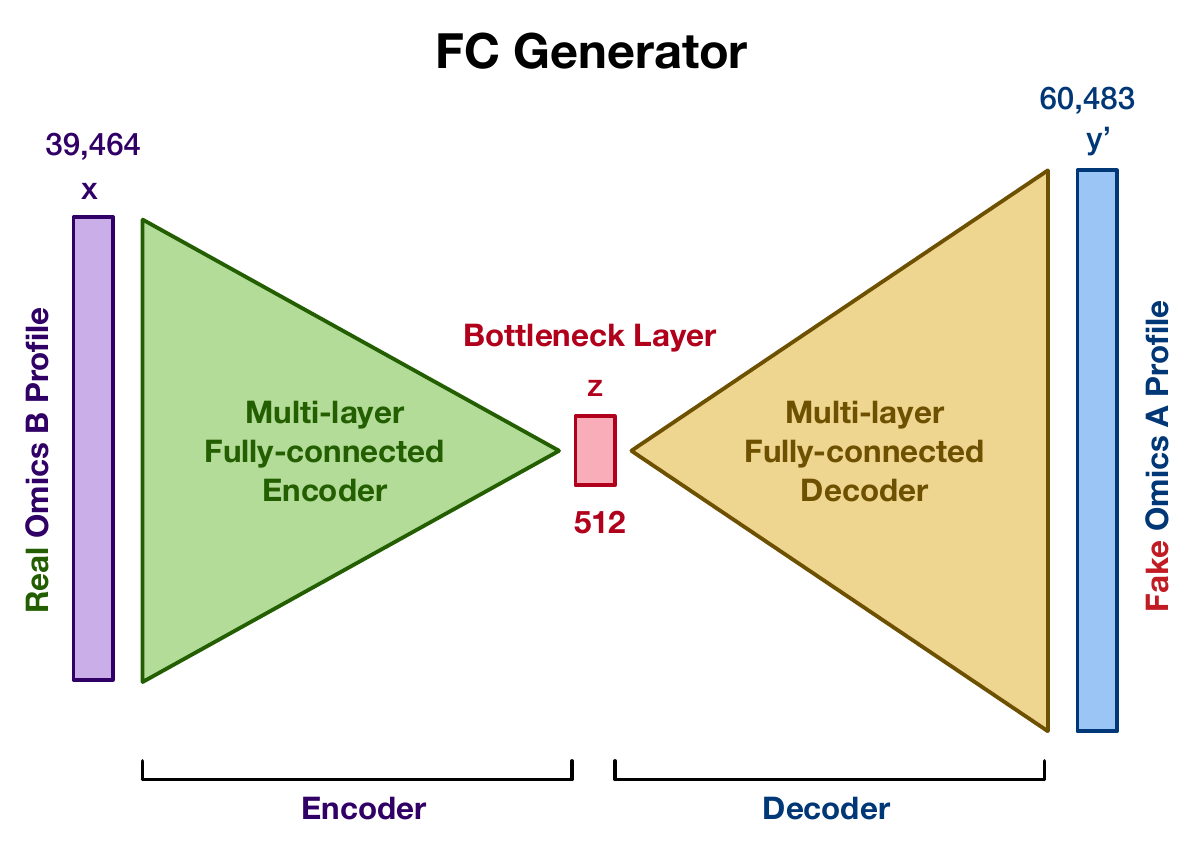}
    \centering
    \caption{The implementation of the OmiTrans generator $G$ using a deep fully-connected (FC) encoder-decoder network. The encoder and decoder of $G$ consist of multiple fully-connected blocks (FC + norm + dropout + activation). The dimensionality of the bottleneck layer is a hyper-parameter normally set to 512. Here we used the omics translation from DNA methylation (omics B) to gene expression (omics A) as an example. The dimensionalities of the two omics types were marked in the diagram.}
    \label{fig:fcg}
\end{figure}

\begin{figure}[t!]
    \includegraphics[width = 0.8\hsize]{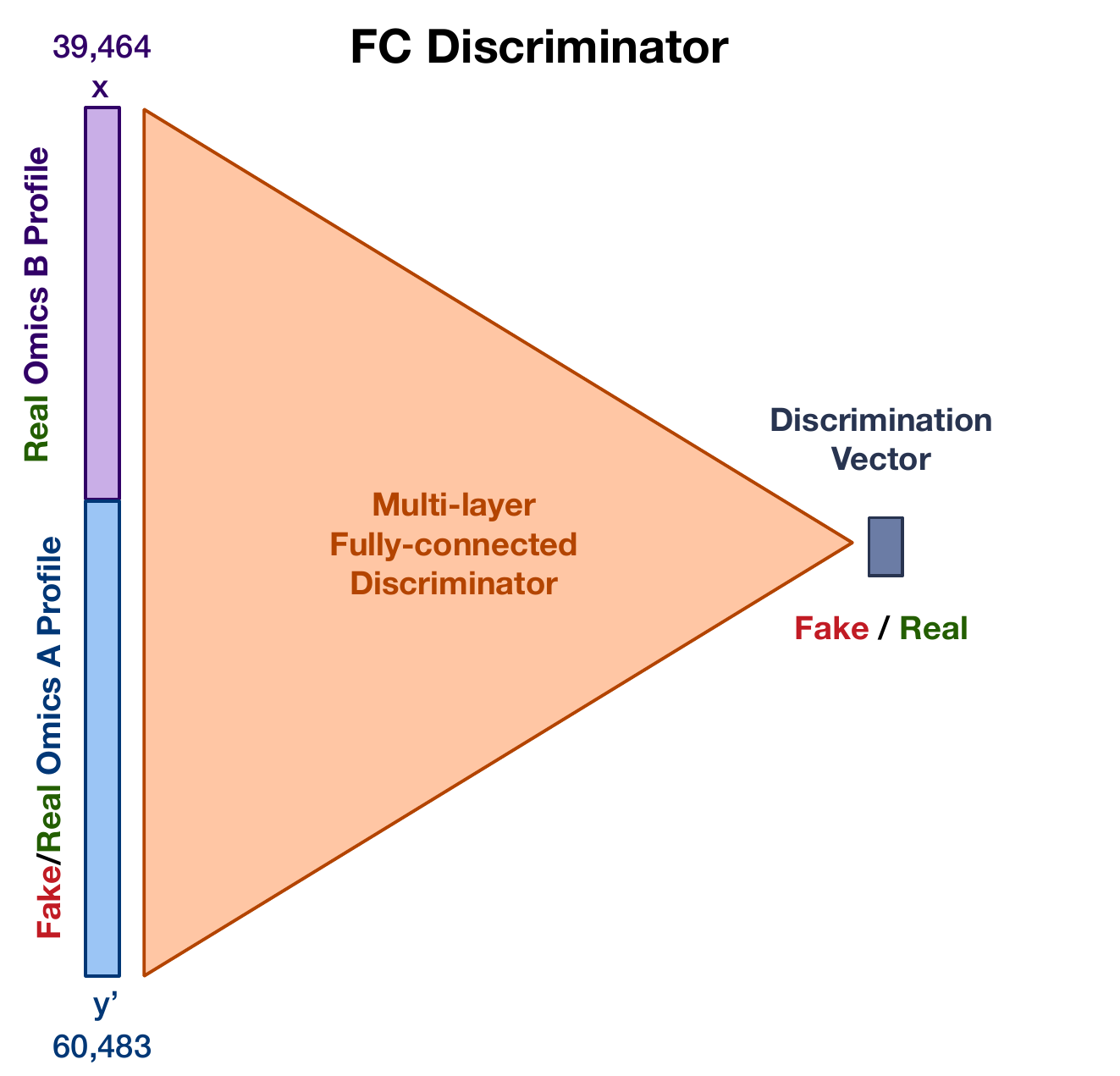}
    \centering
    \caption{The implementation of the OmiTrans discriminator $D$ using a deep fully-connected (FC) discriminator network, which consists of multiple fully-connected blocks (FC + norm + dropout + activation). The output of the discriminator network is a discrimination vector determining whether the input omics A is fake or real. Here we used the omics translation from DNA methylation (omics B) to gene expression (omics A) as an example. The dimensionalities of the two omics types were marked in the diagram.}
    \label{fig:fcd}
\end{figure}

Therefore, the most straightforward implementation of the OmiTrans generator $G$ is an encoder-decoder network using multiple fully-connected (FC) blocks which is comprised of a fully-connected layer, a normalisation layer, a dropout layer and an activation layer, as illustrated in Figure \ref{fig:fcg}. Unlike the translation of image data, the correlation between the features of the input and features of the output is not just one-to-one mapping but a combination of one-to-one, one-to-many, many-to-one and many-to-many mappings. Some of the correlations between features (e.g., gene, CpG site, miRNA, and protein) of two omics types have been discovered by biologists, whereas others are still hidden. Fully-connected layers are able to capture any kind of the aforementioned correlations, which justifies the application of FC blocks in the generator. Similar to the FC generator, the OmiTrans discriminator $D$ can also be implemented by fully-connected blocks as shown in Figure \ref{fig:fcd}. The input vectors of omics A and omics B were first concatenated together and then fed to a multi-layer fully connected network to produce an output vector discriminating whether the input vector of omics A was real or fake.

\begin{figure}[t!]
    \includegraphics[width = 1\hsize]{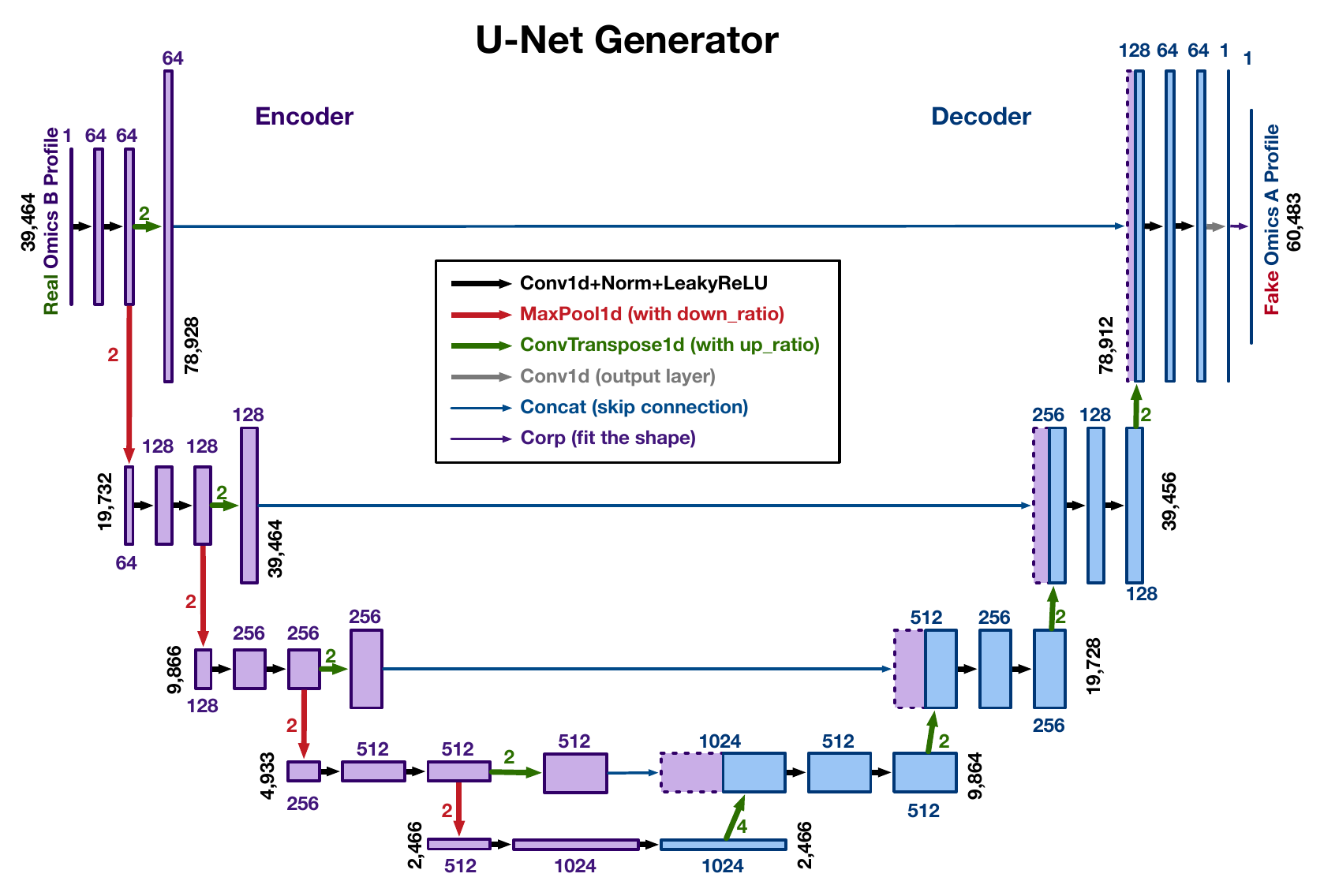}
    \centering
    \caption{The implementation of the OmiTrans generator $G$ using a U-Net network. The channel number and feature number of each tensor in the network were marked in the diagram. Each type of operation group (e.g., conv1D + norm + LeakyReLU) was illustrated using arrows with different colours and thicknesses shown in the legend. Skip connections between each layer $i$ and layer $n-i$ ($n$ stands for the total number of layers in the U-Net) can also be seen in the diagram.}
    \label{fig:unet}
\end{figure}

\begin{figure}[t!]
    \includegraphics[width = 0.85\hsize]{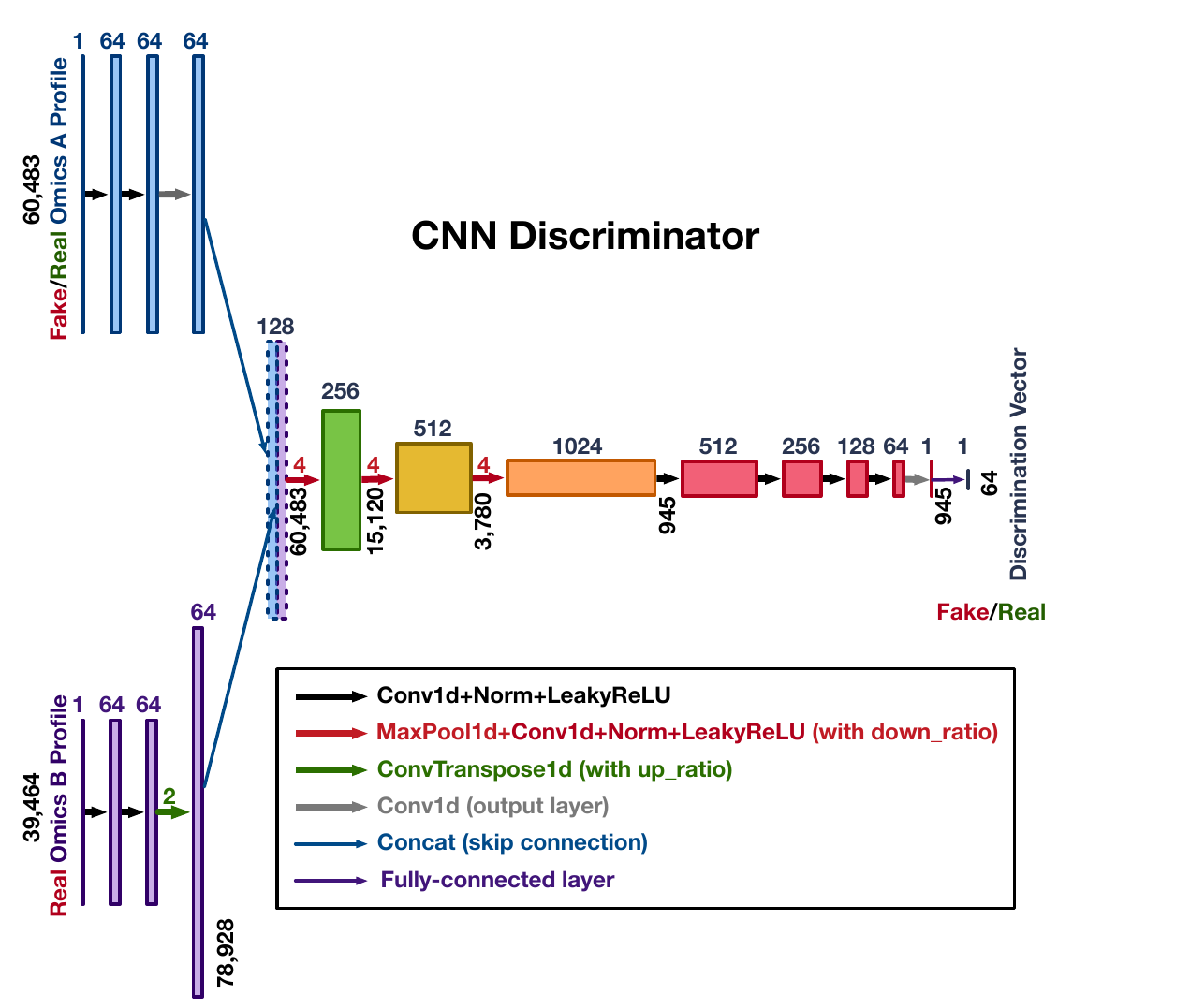}
    \centering
    \caption{The implementation of the OmiTrans discriminator $D$ using a CNN network. The channel number and feature number of each tensor in the network were marked in the diagram. Each type of operation group (e.g., conv1D + norm + LeakyReLU) was illustrated using arrows with different colours and thicknesses shown in the legend. The output of the discriminator network is a discrimination vector determining whether the input omics A is fake or real.}
    \label{fig:convD}
\end{figure}

Although convolutional neural networks are naturally not suitable for omics data, we can still make 1D convolutional layers workable by applying some additional preprocessing steps to the omics data including balancing the dimensionalities of the two omics types and rearranging the orders of the molecular features (e.g., gene, CpG sites, miRNA) based on their genomic location (a.k.a., location alignment). Thus, CNN-based network can be applied for the generator and discriminator of OmiTrans. We modified the U-Net architecture \citep{Ronneberger2015UNetCN} to make it compatible with omics data, as illustrated in Figure \ref{fig:unet}. The discriminator of OmiTrans can also be implemented by a CNN network as shown in Figure \ref{fig:convD}. The performance of both the FC-based architecture and the CNN-based architecture were evaluated in our experiments, and any state-of-the-art or future network can also be added to the OmiTrans framework as the generator or discriminator using the same source code with minimal modification.

\subsection{Comparing Methods}
The OmiTrans framework was built on PyTorch \citep{Paszke2019PyTorchAI}, and has been made open source through GitHub (\url{https://github.com/zhangxiaoyu11/OmiEmbed/}). OmiTrans is compatible with the OmiEmbed \citep{Zhang2021OmiEmbedAU} multi-omics multi-task framework and well-organised with modular code structures, predefined packages and easy-to-follow tutorials. The model of OmiTrans was trained on our experiment platform with two NVIDIA Titan X GPUs, one 6-core 3.40GHz Intel Core i7-6800K CPU, and 96 GB of memory, which is a common configuration that is easy to reproduce.

We have also compared OmiTrans with other methods including the latest TDimpute \citep{Zhou2020ImputingMR} that using transfer learning-based neural networks to impute missing RNA-Seq data from DNA methylation, the traditional linear regression (LR), and LASSO regression \citep{Tibshirani1996RegressionSA}. Unlike the implementation of \cite{Zhong2019PredictingGE} that only connected each gene with CpG probes mapped to it, we fully connected each gene to each CpG site for LR and LASSO to capture all of the potential correlations between genome-wide DNA methylation profile and gene expression profile. TDimpute was ran on the same platform mentioned above for OmiTrans, whereas LR and LASSO were not able to perform on our experiment platform due to the extremely high memory usage and the extremely long running time. Therefore, we ran the experiments of LR and LASSO on Alibaba Cloud Elastic Container Service (ECS) with 80 vCPU of Intel Xeon (Cascade Lake) Platinum 8269CY and 192 GB of memory using multiprocessing and celer \citep{Massias2018CelerAF, massias2020dual}. Even with this top configuration, the training time of LASSO is around 154 hours, which is unaccepted for normal use comparing to the 12-hour training time of OmiTrans.

\section{Results and discussion}
OmiTrans is generative adversarial networks (GANs) based framework designed for omics data translation between any two omics types. Here we used the conversion from genome-wide DNA methylation profile to RNA-Seq gene expression profile as an example to demonstrate the performance of OmiTrans and other comparing methods.

\subsection{Reconstruction Performance}
Since both the DNA methylation data and the corresponding gene expression data are available in the TCGA dataset, we are able to compare the synthetic data with the original data to evaluate the reconstruction performance of each model. The TCGA dataset was randomly separated into training, validation and testing sets. Results on the testing set using the trained model were shown in this section.

As tabulated in Table \ref{tab:recon}, we used nine different metrics to evaluate the reconstruction performance of six methods: FC-based OmiTrans, CNN-based OmiTrans, TDimpute \citep{Zhou2020ImputingMR}, LASSO regression \citep{Tibshirani1996RegressionSA}, conditional GAN, and traditional linear regression (LR). The first four metrics were used to measure the distance between the original omics data and the synthetic omics data, including mean square error (MSE), root mean square error (RMSE), mean absolute error (Mean AE) and median absolute error (Median AE). The remaining five metrics are variants of the coefficient of determination ($R^2$) which is a more intuitively informative metric measures how well observed data are replicated by a model \citep{Chicco2021TheCO}. \cite{Zhong2019PredictingGE} used featurewise $R^2$ ($R^2_f$), while \cite{Zhou2020ImputingMR} used samplewise $R^2$ ($R^2_s$). To represent the results more comprehensively, we included both variants in the evaluation. Since featurewise (i.e., calculate the coefficient of determination for each gene) $R^2$ ($R^2_f$) accords with the original definition of $R^2$ more faithfully, we further analysed $R^2_f$ by calculate both the mean $R^2_f$ and median $R^2_f$, counted the number of genes with $R^2_f$ larger than the threshold, and computed the percentage of genes with $R^2_f$ larger than the threshold. We set threshold to 0.3 in our analysis.

\begin{table*}[t!]
    \caption{Reconstruction performance of six methods using nine different metrics. For the first four metrics, lower value means better performance. For the remaining metrics, higher value means better performance.\label{tab:recon}}
    \tabcolsep=0pt
    \begin{tabular*}{\textwidth}{@{\extracolsep{\fill}}lccccccccc@{\extracolsep{\fill}}}
    \toprule
                   & MSE    & RMSE   & Mean AE & Median AE & Mean $R^2_s$ & Mean $R^2_f$ & Median $R^2_f$ & \# $R^2_f>0.3$\ & \% $R^2_f>0.3$\\ \midrule
    OmiTrans-FC  & \textbf{0.1097} & \textbf{0.3204} & \textbf{0.1538}  & \textbf{0.0396}    & \textbf{0.9453}      & \textbf{0.3556}      & \textbf{0.4167}        & \textbf{37,717}                      & \textbf{62.36\%}                    \\
    OmiTrans-CNN & 0.1747 & 0.4062 & 0.1948  & 0.0453    & 0.9131      & -3.4496     & 0.0506        & 15,145                      & 25.04\%                    \\
    TDimpute       & 0.1141 & 0.3279 & 0.1809  & 0.0872    & 0.9432      & -27.8159    & 0.3210        & 31,097                      & 51.41\%                    \\
    LASSO          & 0.1162 & 0.3295 & 0.1645  & 0.0513    & 0.9422      & 0.3491      & 0.3440        & 32,798                      & 54.23\%                    \\
    Conditional GAN            & 0.1988 & 0.4345 & 0.2035  & 0.0413    & 0.9010      & -0.3598     & 0.0950        & 14,023                      & 23.19\%                    \\
    Linear Regression        & 0.4602 & 0.6666 & 0.5128  & 0.4178    & 0.7701      & -2078.2833  & -3.7674       & 4,922                       & 8.14\% \\
    \botrule                    
    \end{tabular*}
\end{table*}

OmiTrans with the FC-based architecture got the best performance in all nine different metrics, which outperformed the CNN-based OmiTrans with a U-Net generator. This may be caused by the fundamental properties of convolutional neural networks like sparse connectivity and weights sharing that do not fit the format of omics data even with location alignment. The hyper-parameters used for the FC-based OmiTrans were listed in Table \ref{tab:hyper}. Epoch number of the learning rate decay is 100, which means that in the last 100 epochs the learning rate decreased to zero linearly. As for hyper-parameters and network architectures of the comparing method TDimpute, we adopted the exact same configurations as mentioned in their work \citep{Zhou2020ImputingMR}. For conditional GAN, we used objective function (\ref{eq:minmax}) instead of objective function (\ref{eq:finalobj}). The hyper-parameters and network architectures of conditional GAN were the same as FC-based OmiTrans.

\begin{table}[t!]
    \caption{Hyper-parameters used for OmiTrans with the FC-based architecture.\label{tab:hyper}}
    \tabcolsep=0pt
    \begin{tabular*}{\columnwidth}{@{\extracolsep{\fill}}ll@{\extracolsep{\fill}}}
        \toprule
        Hyper-parameter & Value \\ \midrule
        Latent dimension & 256 \\
        Learning rate of the generator $G$& $1\times10^{-4}$ \\
        Learning rate of the discriminator $D$& $1\times10^{-4}$ \\
        Batch size & 128 \\
        Epoch number - total & 800 \\
        Epoch number - decay & 100 \\ 
        Dropout rate & 0.2 \\
        \botrule
    \end{tabular*}
\end{table}

The OmiEmbed \citep{Zhang2021OmiEmbedAU} multi-omics multi-task framework can be used to further compared the performance of each method. First we fully trained an OmiEmbed \citep{Zhang2021OmiEmbedAU} model on the pan-cancer multi-class classification benchmark task mentioned in \cite{Zhang2019IntegratedMA} using our TCGA training set. The pre-trained model was then fed with the original testing data and the six fake testing data synthesised by the aforementioned six methods, and got the classification performance of each method as shown in Table \ref{tab:class}. The better classification performance means the synthetic data is more distinguishable, considering the cancer label of each sample, and more similar to the original data. FC-based OmiTrans still got the best performance in all the five classification metrics, including accuracy, precision, recall, F1 score (F1) and area under the receiver operating characteristic curve (AUC).

\begin{table}[t!]
    \caption{Pan-cancer classification performance on the original testing data and six synthetic testing data.\label{tab:class}}
    \tabcolsep=0pt
    \begin{tabular*}{\columnwidth}{@{\extracolsep{\fill}}lccccc@{\extracolsep{\fill}}}
    \toprule
    & Accuracy & Precision & Recall & F1     & AUC    \\
    \midrule
    Original Data     & \textsl{0.9571}   & \textsl{0.9133}    & \textsl{0.9172} & \textsl{0.9142} & \textsl{0.9973} \\
    \midrule
    OmiTrans-FC       & \textbf{0.9444}   & \textbf{0.8918}    & \textbf{0.8878} & \textbf{0.8863} & \textbf{0.9948} \\
    OmiTrans-CNN      & 0.9130   & 0.8741    & 0.8251 & 0.8257 & 0.9931 \\
    TDimpute          & 0.9356   & 0.8867    & 0.8626 & 0.8580 & 0.9946 \\
    LASSO             & 0.9307   & 0.8865    & 0.8688 & 0.8622 & 0.9927 \\
    Conditional GAN   & 0.8745   & 0.7812    & 0.8007 & 0.7839 & 0.9873 \\
    Linear Regression & 0.8921   & 0.8471    & 0.7922 & 0.7965 & 0.9895 \\
    \botrule                    
    \end{tabular*}
\end{table}

\begin{figure}[t!]
    \includegraphics[width = 1\hsize]{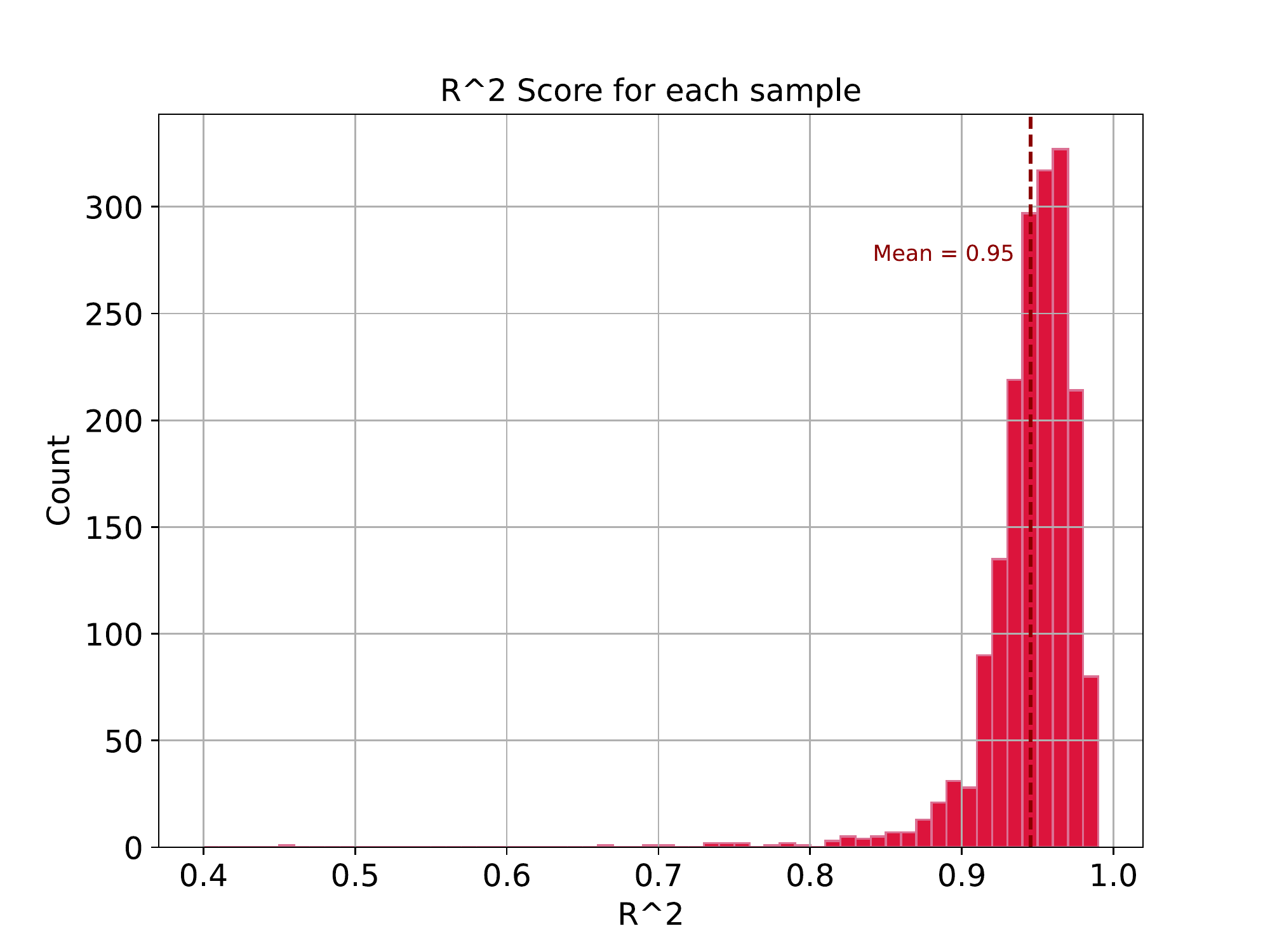}
    \centering
    \caption{Histogram of the samplewise $R^2$ scores ($R^2_s$).}
    \label{fig:hist_sample}
\end{figure}

\begin{figure}[t!]
    \includegraphics[width = 1\hsize]{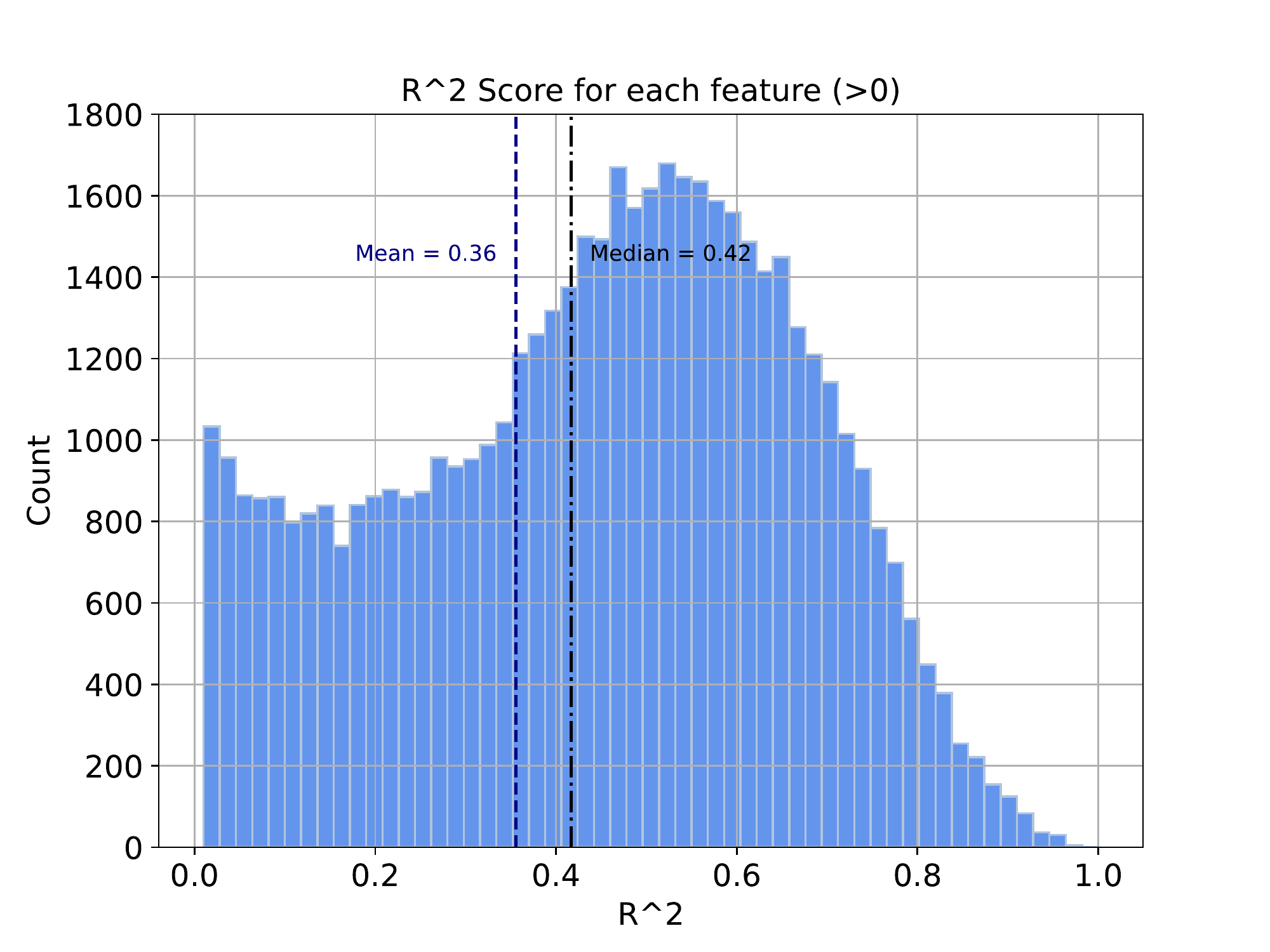}
    \centering
    \caption{Histogram of the featurewise $R^2$ scores ($R^2_f$) that larger than zero.}
    \label{fig:hist_feature}
\end{figure}

\begin{figure*}[t!]
    \includegraphics[width = 1\hsize]{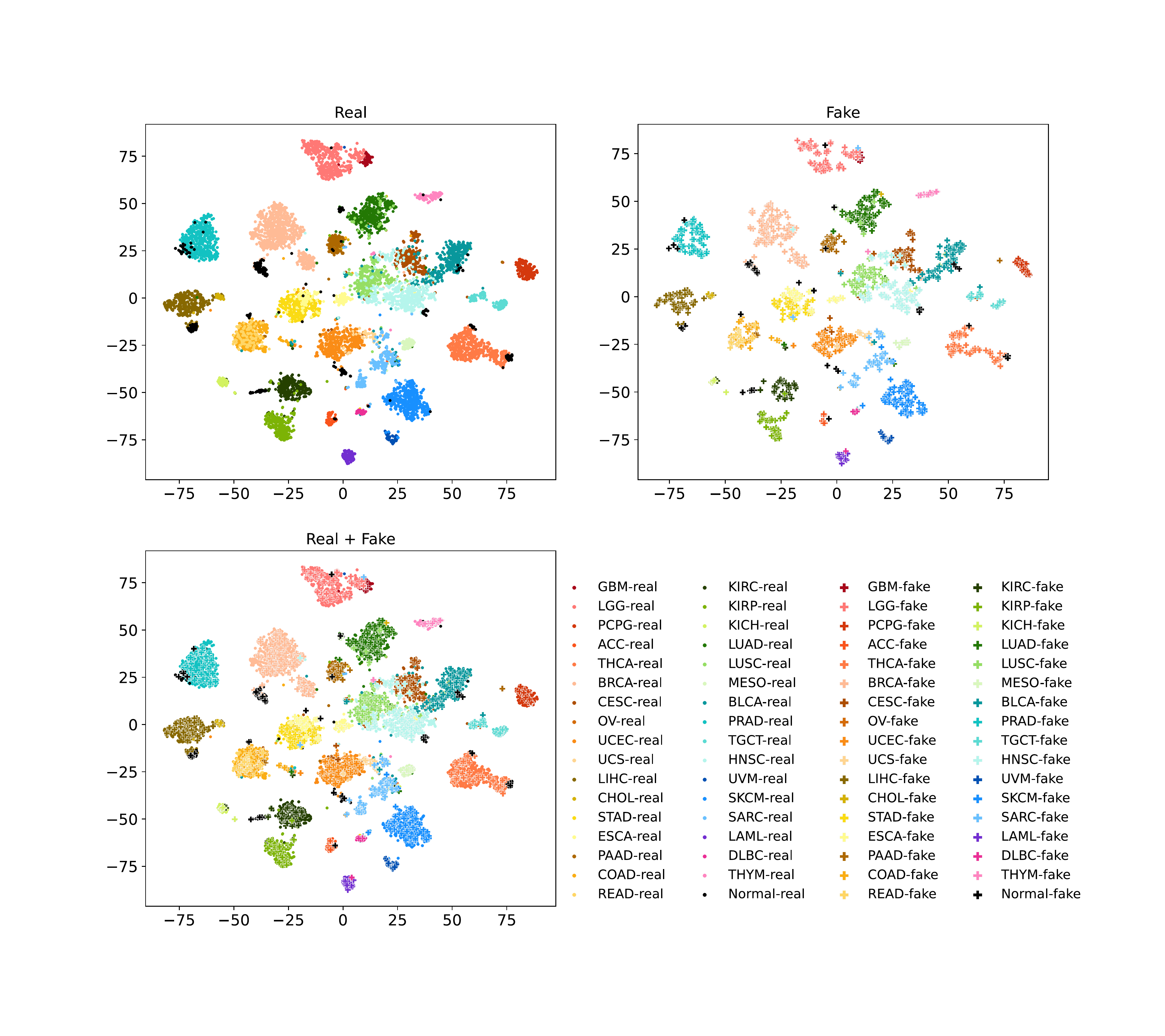}
    \centering
    \caption{The scatter graphs of the original training set of TCGA, the fake testing set of TCGA synthesised by OmiTrans, and both sets together. Samples with different tumour types were marked with different colours, original data and synthetic data were marked with different symbols, as shown in the legend.}
    \label{fig:latent_real_fake}
\end{figure*}

Since the FC-based OmiTrans got the best performance of reconstructing genome-wide gene expression profile from DNA methylation data, we further analysed the results got from OmiTrans (henceforth OmiTrans refers to the FC-based OmiTrans). The histogram of the samplewise $R^2$ and the featurewise $R^2$ were illustrated in Figure \ref{fig:hist_sample} and Figure \ref{fig:hist_feature}. The mean and median $R^2$ values were also marked in the histogram. To observe the distribution of the gene expression data synthesised by OmiTrans visually, we used OmiEmbed \citep{Zhang2021OmiEmbedAU} to reduce the dimensionalities of the synthetic testing set and the original training set of TCGA to 128 and then used t-SNE \citep{Maaten2008VisualizingDU} to visualise both the synthetic and original data in scatter graphs as shown in Figure \ref{fig:latent_real_fake}. As can be seen in in Figure \ref{fig:latent_real_fake}, the synthetic data shared the similar distribution in the latent space, which indicated that OmiTrans faithfully reconstructed the gene expression profiles from the DNA methylation data.

\subsection{Testing on Individual Dataset}
\begin{figure*}[t!]
    \includegraphics[width = 1\hsize]{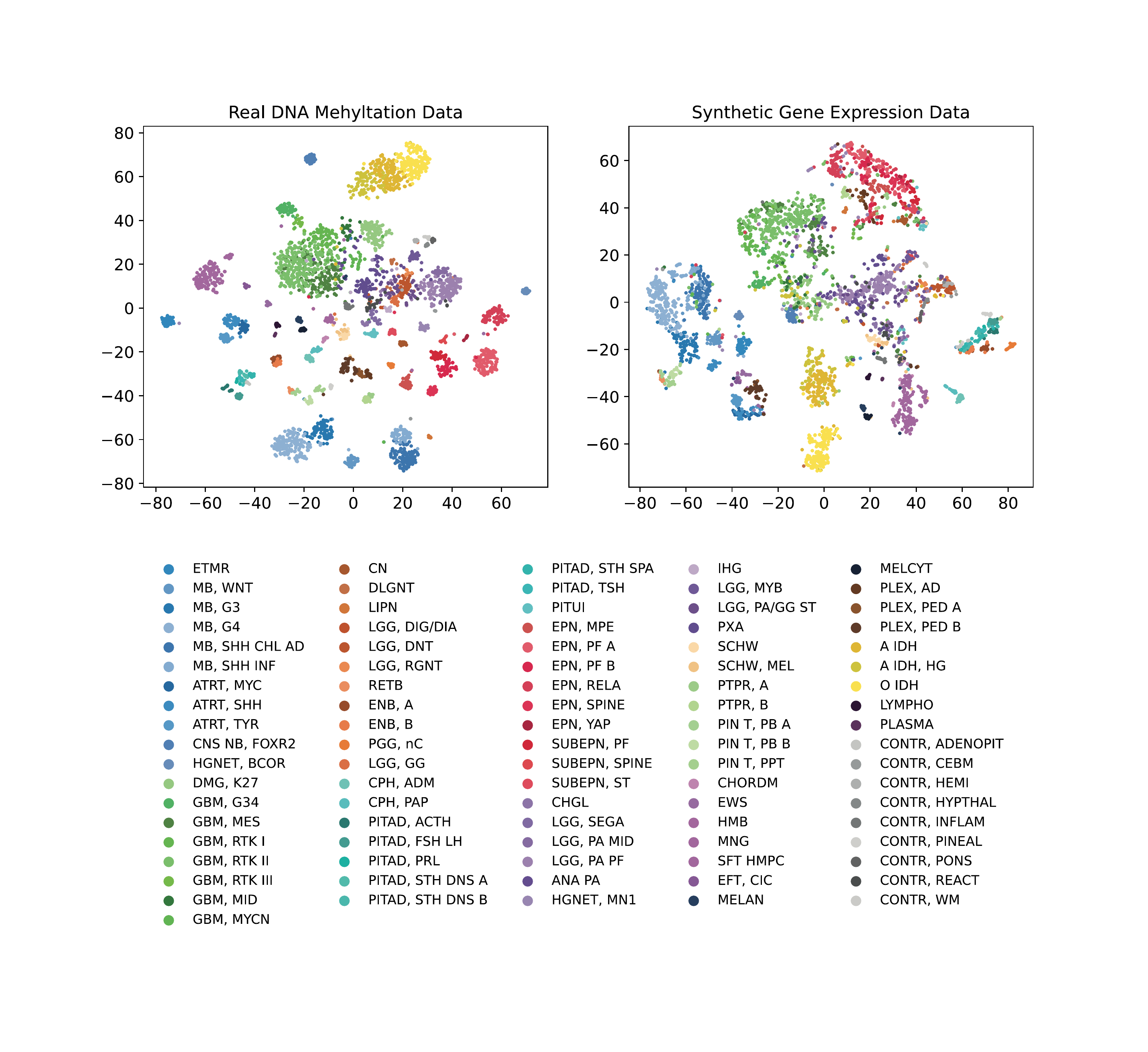}
    \centering
    \caption{The scatter graphs of the original DNA methylation data of the BTM dataset, the fake gene expression data of BTM synthesised by a OmiTrans generator trained on the TCGA dataset. Samples with different brain tumour types or control regions were marked with different colours, brain tumour types and control regions belonging to the same upper-level class were marked in similar colours, as shown in the legend.}
    \label{fig:latent_brain_real_fake}
\end{figure*}

All of the experiments aforementioned were tested on TCGA. Thus, it remained unknown whether a OmiTrans model trained on one dataset still worked on other individual dataset. To further evaluated the generalisation of OmiTrans, we applied the generator of an OmiTrans model that had been fully trained on the TCGA dataset to a new and previously unseen dataset: the brain tumour DNA methylation dataset (BTM) GSE109381 \citep{Capper2018DNAMC}. The OmiTrans generator synthesised gene expression profile from each DNA methylation profile in the BTM dataset. We used OmiEmbed \citep{Zhang2021OmiEmbedAU} to reduce the dimensionalities of the original DNA methylation data and the synthetic gene expression of BTM to 128 and then used t-SNE \citep{Maaten2008VisualizingDU} to visualise them, as illustrated in Figure \ref{fig:latent_brain_real_fake}.

Even though the OmiTrans model was trained on the TCGA dataset and had never seen any data from the BTM dataset, it was able to synthesise gene expression profile that we did not have for every sample in BTM. As we can see in Figure \ref{fig:latent_brain_real_fake}, the synthetic gene expression data kept the multi-level hierarchical clustering pattern of the original DNA methylation data, although the OmiTrans model had zero information about the BTM dataset and the brain tumour classification criteria, which means a pre-trained OmiTrans model can be easily transferred to another dataset without any a priori knowledge. Unlike TCGA and BTM dataset, most omics datasets are small-scale which is suitable for deep learning. However, with the model generalisation of OmiTrans, we can pre-trained a OmiTrans model on large multi-omics dataset like TCGA and adopted the model to the dataset-of-interest with small sample size, which largely increase the applicability and practicability of OmiTrans.

\section{Conclusion}
Omics-to-omics translation is a brand new research area that has been hardly studied. Here we proposed OmiTrans which is the first GANs based omics-to-omics translation framework to the best of our knowledge. OmiTrans is able to faithfully generate synthetic omics data from corresponding original data of anther omics type. With the model generalisation ability, a OmiTrans model can be trained on a large-scale multi-omics dataset like TCGA and adopted to small-scale dataset-of-interest, which largely increase the applicability and practicability.


\section{Key Points}
\begin{itemize}
    \item OmiTrans is the first GANs-based omics-to-omics translation framework.
    \item OmiTrans is able to faithfully generate synthetic omics data from corresponding original data of anther omics type.
    \item OmiTrans outperformed current state-of-the-art methods.
    \item OmiTrans has great applicability, practicability, and model generalisation.
\end{itemize}

\section{Availability}
The source code have been made publicly available on GitHub \url{https://github.com/zhangxiaoyu11/OmiTrans/}. The TCGA dataset can be downloaded from the UCSC Xena data portal \url{https://xenabrowser.net/datapages/}. The release of OmiTrans has also been stored in Zenodo under the doi:10.5281/zenodo.5728496 (\url{https://zenodo.org/record/5728496}). The BTM dataset is available from GEO (\url{https://www.ncbi.nlm. nih.gov/geo/query/acc.cgi?acc=GSE109381}) with the accession ID GSE109381. 

\section{Acknowledgments}
This work was supported by the European Union's Horizon 2020 research and innovation programme under the Marie Sk\l{}odowska-Curie grant agreement [764281].

\section{Competing interests}
The authors declare that they have no conflict of interest.

\bibliographystyle{abbrvnat}
\bibliography{reference}



\begin{biography}{}{\author{Xiaoyu Zhang} is currently a PhD candidate at Data Science Institute, Imperial College London, London, UK.}
\end{biography}

\begin{biography}{}{\author{Yike Guo} is currently the co-director of Data Science Institute, Imperial College London, London, UK and vice-president of Hong Kong Baptist University, Hong Kong, China.}
\end{biography}

\end{document}